\documentclass{jnmp}

\usepackage{amsmath}
\usepackage{amssymb}
\usepackage{amsfonts}
\usepackage{graphicx}

\JNMPnumberwithin{equation}{section}

\newtheorem{theorem}{Theorem}
\newtheorem{lemma}{Lemma}

\theoremstyle{definition}

\newtheorem{conjecture}{Conjecture}

\def\={\stackrel{\mathrm{def}}{=}}
\def\B{\({\cal B}\)}
\def\Bf#1#2#3#4{\begin{array}{rll} u_t=& #1 u_{#3} + #4 \\ v_t=& #2 v_{#3}
\end{array}}
\def\vect#1#2{\left(\begin{array}{c} {#1}\\ {#2}\end{array} \right) }
\def\mat#1#2#3#4{\left( \begin{array}{cc} #1 & #2 \\ #3 & #4 \end{array}
\right)}
\def\C{\mathbb{C}}
\def\Z{\mathbb{Z}}

\begin{document}
\thispagestyle{empty}

\FirstPageHead{8}{4}{2001}{\pageref{firstpage}--\pageref{lastpage}}{Article}

\copyrightnote{2001}{P H van der Kamp and J A Sanders}

\Name{On Testing Integrability}

\label{firstpage}

\Author{Peter H VAN DER KAMP and Jan A SANDERS}

\Address{
Vrije Universiteit,
Faculty of Sciences,
Division of Mathematics \& Computer Science,
De Boelelaan 1081a,
1081 HV Amsterdam,
The Netherlands\\
~~E-mail: peter@cs.vu.nl, jansa@cs.vu.nl}

\Date{Received June 19, 2001;
Accepted July 27, 2001}
\bibliographystyle{plain}
\begin{abstract}
We demonstrate, using the symbolic method together with p-adic and
resultant methods,
the existence of systems with exactly one or two generalized symmetries.
Since the existence of one or two symmetries is often taken as a sure sign
(or as the definition) of
integrability, that is, the existence of symmetries on infinitely many orders,
this shows that such practice is devoid of any mathematical foundation.
Extensive computations show that systems with one symmetry are rather common,
and with two symmetries are fairly rare, at least within the class we have
been considering in this paper.
\end{abstract}
\section{Introduction}
In 1980 an observation was made at least twice by different authors. In
\cite{Fok80} it is
written
\begin{quote}
Another interesting fact regarding the symmetry structure of evolution
equations is that in all
known cases the existence of one generalized symmetry implies the existence
of infinitely many.
\end{quote}
and in \cite{MR82f:58087} the same statement is made together with the footnote
\begin{quote}
This is not true for systems of equations. For example,  the system
\(u_t=u_2+v^2/2, v_t=2v_2\) has a nontrivial group,
but this group is exhausted by the one-parameter (with parameter \(\tau\))
group of transformations: \(u_\tau=u_3+3vv_1, v_\tau=4v_3\).
\end{quote}
Here \(v_2\) stands for \(\frac{\partial^2 v}{\partial x^2}\); we use the
same notation
in this paper.
However, the 'counter-example' given there is an integrable system, cf
\cite{Ba91}, section 3.
In spite of this fact
Fokas adapted the remark and formulated the following conjecture in 1987,
\cite{Fok87}.
\begin{conjecture}[Fokas]
If a scalar equation possesses at least one time-indepen\-dent non-Lie
point symmetry,
then it possesses infinitely many. Similarly for n-component equations one
needs n symmetries.
\end{conjecture}
Besides their mathematical interest, the observation and conjecture are
of some practical importance since they are used to argue that it is enough
to find only one or two symmetries of a system in order to declare it
integrable,
cf \cite{MR99c:58077} and the discussion in \cite{MR2001a:37100}.
This would be reasonable practice if it was simply not possible to prove
integrability, but the methods employed in
\cite{MR99g:35058,MR99i:35005,MR1779258,MR1781148,SW01a} showhow one can
effectively
obtain integrability proofs.
Moreover, four years later Bakirov \cite{Ba91} published the first example
of a non-integrable equation
in the possession of a generalized symmetry. The system
\[
\Bf{5}{}4{v_0^2}
\]
has a sixth order symmetry
\[
\Bf{11}{}{6}{5 v_0v_2 + 4 v_1^2}
\]
as one can easily check.
It was shown (with extensive computer algebra computations) that
there are no other symmetries up to order 53.  The authors of
\cite{MR99i:35005} proved in
1998 that the system of Bakirov does not possess another symmetry at any
higher order, thereby
proving that indeed one symmetry does not imply integrability.
In \cite{KS99} it is proved that there are in fact infinitely many fourth
order systems
with finitely many symmetries. The method used there could be used for
seventh order systems
as well and a system with two symmetries was found, a counterexample to
Fokas' conjecture.
However for systems whose order is more than seven the approach can no
longer be used.
The method introduced here (using resultants) makes it possible to explore
the symmetries of
higher order systems.
\section{The symmetry condition}
We study symmetries of
\[
\Bf{a_1}{a_2}{n}{K(v,v_1,\ldots,v_{n-1})}
\]
where \(a_1,a_2 \in \C\) and \(K\) is polynomial in \(v(x,t)\) and its
derivatives
\(v_k\). We call these {\bf \B--systems}, where \(n\) is the
{\bf order} of the system.

The right hand side of the equation can be interpreted as an element in a
Lie algebra,
the Lie bracket is computed using Fr\'echet derivatives as follows, see
\cite{MR94g:58260}, \cite{Mag78}
\begin{equation}\label{ek}.
\begin{array}[b]{l}
\left[ \vect{a_1 u_n + K}{a_2 v_n} , \vect{b_1 u_m + S}{b_2 v_m} \right] \\ \\
=\mat{b_1 D^m}{D_S}0{b_2 D^m} \vect{a_1 u_n + K}{a_2 v_n} -
\mat{a_1 D^n}{D_K}0{a_2 D^n} \vect{b_1 u_m + S}{b_2 v_m}  \\ \\
=\vect{a_1 D^n S - a_2 D_S v_n - b_1 D^m K + b_2 D_K v_m}{0}.
\end{array}
\end{equation}
We call \(S\) a symmetry of \(K\) if \([K,S]\) vanishes.
\(K\) is called {\bf integrable} when there exist symmetries
on infinitely many orders and {\bf almost integrable} when
there exist symmetries
on finitely many orders, cf. \cite{KS99}.

How to solve the equation \([K,S]=0\), given the order of \(K\) and \(S\)?
First of all, we see that if \(K\) is polynomial, \(S\) has to be
polynomial too, cf. \cite{Ba91}.
This enables us to use the symbolic calculus as developed in
\cite{MR58:22746}. With the symbolic calculus the
equation becomes polynomial and leads to divisibility conditions of
certain elementary polynomials. These can be solved for infinitely many
orders at
once. The necessary and sufficient equations for the ratio of eigenvalues
are obtained directly
without having to specify the nonlinear part explicitly.

Assume that \(K\) and \(S\) are quadratic.
A quadratic differential monomial is transformed into a symmetric
polynomial in two symbols as follows
\[
v_i v_j = \frac{\xi_1^i\xi_2^j+\xi_1^j\xi_2^i}{2}
\]
The expression is symmetrized and divided by the number of
symbol-permutations in order to ensure
that
\[
v_i v_j = v_j v_i
\]
This procedure turns the operation of differentiation into ordinary
multiplication
\[
\begin{array}{ll}
D v_iv_j &= v_{i+1} v_j + v_i v_{j+1} \\
	 &=
\frac{\xi_1^{i+1}\xi_2^j+\xi_1^j\xi_2^{i+1}+\xi_1^i\xi_2^{j+1}+\xi_1^{j+1}\xi_2^
i}{2} \\
	 &= ( \xi_1+\xi_2 )( \frac{\xi_1^i\xi_2^j+\xi_1^j\xi_2^i}{2} ),
\end{array}
\]
like the action of the Fr\'echet derivative on a linear term
\[
\begin{array}{ll}
D_{v_i v_j} v_k &= v_{i+k} v_j + v_i v_{j+k} \\
	        &=
\frac{\xi_1^{i+k}\xi_2^j+\xi_1^j\xi_2^{i+k}+\xi_1^i\xi_2^{j+k}+\xi_1^{j+k}\xi_2^
i}{2} \\
	        &= ( \xi^k_1+\xi^k_2 )(
\frac{\xi_1^i\xi_2^j+\xi_1^j\xi_2^i}{2} ).
\end{array}
\]
The symmetry condition for quadratic polynomials (\(K,S \in \C
[\xi_1,\xi_2]\)) reads
\[
G_n[a_1,a_2] S = G_m[b_1,b_2] K
\]
with the \(G\)--functions
\[
G_n[a_1,a_2](\xi_1,\xi_2) \= a_1(\xi_1+\xi_2)^n-a_2(\xi^n_1+\xi^n_2)
\]
If \(G_m[b_1,b_2] K\) is divisible by \(G_n[a_1,a_2]\) we have a symmetric
polynomial expression for \(S\) which can be
transformed back. Because the \(\xi_1\)-degree of \(K\) (the maximal number
of x-derivatives of \(u\) in the terms of \(K\)) is
smaller than \(n\), the function \(G_n[a_1,a_2]\) cannot divide \(K\).
Therefore \(G_n[a_1,a_2]\) should have a common factor
with \(G_m[b_1,b_2]\) for there being a symmetry.
Suppose we can find \(a_1,a_2,b_1,b_2 \in \C\) such that (\(F,L,T \in \C
[\xi_1,\xi_2]\))
\[
\begin{array}{l}
G_n[a_1,a_2]=FL\\
G_m[b_1,b_2]=FT.
\end{array}
\]
Then the Lie bracket (\ref{ek}) vanishes if one takes \(K=LM\) and
\(S=MT\). One is free to choose \(M
\in \C[\xi_1,\xi_2]\) as long as the \(\xi_1\)-degree of \(K\) remains
smaller than \(n\).
If this sounds a bit too easy, the reader should note that this determines
the system and its symmetry at the same time, and does not say anything about
hierarchies of symmetries.
\section{The use of resultants}
If the resultant of two polynomials vanishes, then their greatest common
divisor has
positive degree.
\begin{lemma}
Second order \B--systems have symmetries at all orders. The ratio of
eigenvalues
(and quadratic part) of the symmetries are fixed.
\end{lemma}
\begin{proof}
Take \(a_1 \neq a_2\) again and \(r \neq -1\). The \(G\)--function is
\[
G_2[a_1,a_2]=\frac{a_1-a_2}{r}(\xi_1-r\xi_2)(r\xi_1-\xi_2) \text{ with }
(a_2-a_1)r^2-2a_1 r-(a_2-a_1)=0,
\]
and the resultant of \((\xi_1-r\xi_2)\) with \(G_m[b_1,b_2]\) vanishes when
\[
\frac{b_1}{b_2}=\frac{1+r^m}{(1+r)^m}.
\]
With this ratio \((r\xi_1-\xi_2)\) is a factor as well because the
\(G\)--function is symmetric
in \(\xi_1,\xi_2\) (the fraction \(\frac{1+r^m}{(1+r)^m}\) is invariant under
\(r\rightarrow \frac{1}{r}\)).
\end{proof}
This implies that to cover systems with finitely many symmetries the degree
of the
common factor of the \(G\)-functions should be higher than 2. Degree 3 is
not enough
because a third degree symmetric polynomial always contains the factor
\(\xi_1+\xi_2\).
Their corresponding systems are always in a hierarchy of first, second or
third order.

We look at factors of degree 4.
\begin{lemma}\label{lem2}
The function \(G_n[1+r^n,(1+r)^n](\xi_1,\xi_2)\)
has a factor of the form
\[
(\xi_1-r\xi_2)(r\xi_1-\xi_2)(\xi_1-s\xi_2)(s\xi_1-\xi_2)
\]
whenever \(U_n\=G_n[1+r^n,(1+r)^n](1,s)=0\).
\end{lemma}
\begin{proof}
The condition \(U_n=0\) is expressing the fact that the ratio of
eigenvalues of the \(G\)-function
containing root \(r\) equals the ratio of eigenvalues of the \(G\)-function
containing root \(s\).
\end{proof}
In the following we disregard the trivial factors of \(U_n\) which are
\((r-s)(rs-1)\) for all
\(m\) and \((r+1)(s+1)\) when \(m\) is odd.
\begin{lemma} Take \(n>3.\) To obtain all eigenvalues of \(n^{\text{th}}\)
order \B-systems with
a symmetry on order \(m\) one calculates the resultant of \(U_n\) and
\(U_m\) with respect to \(s\)
and applies the map \(r\rightarrow\frac{1+r^m}{(1+r)^m}\) to its roots.
\end{lemma}
\begin{proof}
If the resultant of \(U_n\) and \(U_m\) vanishes for some \(r \in \C\) then
by the
previous Lemma \(G_n[1+r^n,(1+r)^n](\xi_1,\xi_2)\) and
\(G_m[1+r^m,(1+r)^m](\xi_1,\xi_2)\)
have a common fourth order factor. This implies that the \(n^{\text{th}}\)
order \B--system with
eigenvalues \(a_1=1+r^n,a_2=(1+r)^n\) and quadratic part \(G_n[a_1,a_2]\)
divided by this
fourth order factor has a symmetry on order \(m\).
\end{proof}

Example of Bakirov: The resultant of \(U_4\) and \(U_6\) with respect to
\(s\) contains the factor
\[
f(r)=2r^4+10r^3+15r^2+10r+2.
\]
We have that
\[
1+r^4 \text{ mod } f(r) \equiv 5\frac{r(2r^2+2+3r)}{-2}
\]
and
\[
(1+r)^4 \text{ mod } f(r) \equiv \frac{r(2r^2+2+3r)}{-2}.
\]
Their ratio is \(5\), the ratio of the eigenvalues of the Bakirov system.
As expected \(G_4[5,1](1,r)\) is proportional to \(f(r)\).
\section{The use of p-adic numbers}
The use of p-adic methods in integrability theory was initiated in
\cite{MR99i:35005}.
For an introduction in p-adic number theory, see \cite{MR98h:11155}.
In this section we give a more expanded proof of the fact that the Bakirov
system
contains exactly one symmetry.

The p-adic field is notated by \(\Z_p\) where \(p\) is some prime number.
Its elements
are represented by series of the form \(\sum_{n \geq 0} a_n p^n\)
with coefficients \(a_n\in\Z/p\). The p-adic
expansion of an positive integer is just its base \(p\) representation. For
rational numbers we
can get an infinite sequence. Examples: in \(\Z_5\) we have
\[
\begin{array}{l}
57=2\cdot5^0+1\cdot5^1+2\cdot5^2 \\
\frac{3}{4}=1+\frac{1}{1-5}=2\cdot5^0+\sum_{i=1}1\cdot5^i.
\end{array}
\]
An element is invertible (in \(\Z_p^{\times}\)) if
it is nonzero modulo \(p\), that is: \(a_0\neq 0\).
\subsection{Hensels Lemma}
The following lemma gives a method to check whether
a polynomial has a root in \(\Z_p^\times\).
\begin{lemma}[Hensel]
A polynomial
\[
f(r)=\sum_{i=0}^m c_ir^i \text{ with } c_i \in \Z_p
\]
has a root in \(\Z_p^\times\) if there exists an \(\alpha_1 \in \Z/p\) such
that
\begin{itemize}
\item \(f(\alpha_1) \equiv 0 \text{ mod } p\)
\item \(\frac{df}{dr}(\alpha_1) \not \equiv 0 \text{ mod } p\).
\end{itemize}
\end{lemma}
\begin{proof} It is possible to construct a sequence \{\(\alpha_n\)\} with
\[
\begin{array}{l}
\alpha_n \text{ mod } p^{n-1} \alpha_{n-1} \\
f(\alpha_n) \equiv 0 \text{ mod } p^n
\end{array}
\]
Calculate \(\beta \in \Z/p\) such that
\[
0=f(\alpha_{n+1})=f(\alpha_n+\beta p^n) \equiv f(\alpha_n) +
\frac{df}{dr}(\alpha_1)\beta p^n \text{ mod } p^{n+1} .
\]
By the induction hypotheses there exists a \( \gamma \in \Z/p\) such that
\[
f(\alpha_n) \equiv \gamma p^n \text{ mod } p^{n+1}
\]
Substituting this and dividing by \(p^n\) gives an equation that can be solved
in \(\Z/p\):
\[
\beta \equiv -\gamma(\frac{df}{dr}(\alpha_1))^{-1} \text { mod } p.
\]
Since the first step of the induction is part of the hypotheses,
this concludes the proof.
\end{proof}
For example the square roots of 2 are in \(\Z_7\). Take
\[
f(r)=r^2-2
\]
Then
\[
\begin{array}{ll}
f(3)\equiv 0 \text{ mod } 7 , & f(4) \equiv 0 \text{ mod } 7 \\
\frac{df}{dr}(3) \equiv 6 \text{ mod } 7 , & \frac{df}{dr}(4) \equiv 1
\text{ mod } 7
\end{array}
\]
So Hensels Lemma can be applied. The number 3 is lifted as follows. Modulo
\(7^2\) we have
\[
f(3)=1\cdot7
\]
so \(\gamma=1\).  The inverse of 6 in \(\Z_7\) is 6. Then \(\beta \equiv
-1\cdot6 \equiv 1\).
Indeed
\[
f(3+1\cdot7)=2\cdot7^2
\]
One step further gives
\[
f(3+1\cdot7+2\cdot7^2)=6\cdot7^3+4\cdot7^4
\]
This shows that the method of Hensel is constructive.
\subsection{The method of Skolem}
Skolems method allows us to conclude that there exist only a finite number
of symmetries.
At first sight it looks a bit technical, but it is extremely powerful in
our context.
The method is based on the fact that if some equation does not have a
solution in some p-adic
field then it can not have a solution in \(\C\). Moreover the method
reduces the number
of orders that need to be checked to a finite number.

If \(x_i \in \Z_p^\times\) then by Fermats little theorem there exists a \( y_i
\in \Z_p\) such that \(x_i^{p-1}=1+y_ip\). Let
\[
u^m_n\=\sum_{i=1}^jc_iy_i^mx_i^n
\]
For instance, \(U_n\), as defined in lemma \ref{lem2}, has the form of
\(u^0_n\) with
\(c_i=(-1)^i\) and \(j=4\).
\begin{lemma}[Skolem] \label{skol1}
If \(u^0_k \not\equiv 0\) mod \(p\) then \(\forall r,\ u^0_{k+r(p-1)} \neq 0\).
\end{lemma}
\begin{proof}
\[
u^0_{k+r(p-1)} = \sum_{i=1}^jc_ix_i^k(1+y_ip)^r \equiv u^0_k \text{ mod }
p\neq 0.
\]
Therefore \(u^0_{k+r(p-1)}\) itself is \(\neq 0\).
\end{proof}
\begin{lemma}[Skolem] \label{skol2}
If \(u_k^0=0\) and \(u^1_k\not\equiv 0\) mod \(p\) then \(\forall r>0\) we
have \(u^0_{k+r(p-1)} \neq 0\).
\end{lemma}
\begin{proof}
Assume \(u^0_{k+r(p-1)} = 0\)
\[
0 = \sum_{i=1}^j c_ix_i^k(1+y_ip)^r = \sum_{t=1}^r \left( \begin{array}{c}
r \\ t
\end{array} \right) p^tu^t_k
\]
use
\[
\frac{1}{r} \left( \begin{array}{c} r \\ t \end{array} \right) =
\frac{1}{t} \left( \begin{array}{c} r-1 \\ t-1\end{array} \right)
\]
and divide by \(pr\) to get
\[
u^1_k + \sum_{t=2}^r \left( \begin{array}{c} r-1 \\ t
-1\end{array} \right) \frac{p^{t-1}}{t} u^t_k = 0
\]
This contradicts the second assumption since \(\frac{p^{t-1}}{t}\) always
contains a factor \(p\).
To see this write \(t=p^\alpha s\) with \(p\not|\ s\). Then \(s\) is
invertible and
\[
\frac{p^{t-1}}{t}=\frac{1}{s}p^{p^\alpha s-\alpha-1}
\]
The power of \(p\) is bigger than 1 for when \(\alpha=0\) we know \(s \geq
2\) and when \(\alpha \neq 0\) we have
\(s \geq 1\) and \(p^\alpha \geq \alpha +2\) (because \(p > 2\)).
Hence we conclude \(u^0_{k+r(p-1)} \neq 0\).
\end{proof}
With the lemmas of Skolem one has to search a prime number \(p\) such that the
\(x_i\) are  in the field \(\Z_p^\times\), and check the conditions for
finitely many orders
(\(p\)-2). The computations one has to do are all modulo \(p\) or \(p^2\).
\section{The Bakirov system}
Here is how to use these lemmas for the Bakirov system. We let \(p\)
increase and look for p-adic
roots of the resultant \(2r^4+10r^3+15r^2+10r+2\). The first prime such
that all conditions are satisfied is 181.
In \(\Z/181\) we find \(f(66)=f(139)=0\). These numbers can be lifted to
elements of
\(\Z^{\times}_{181}\). Modulo \(p^2\) they are
\[
r \equiv 66 + 13 p,\ s \equiv 139 + 29 p
\]
The function \(U_m(r,s)\) is has the form \(u^0_m\) with \(c_i=(-1)^i\),
\(j=4\) and
\[
\begin{array}{l}
x_1=1+s \equiv 140 + 29 p \text{ mod } p^2 \\
x_2=1+r \equiv 67 + 13 p \text{ mod } p^2 \\
x_3=r(1+s) \equiv 9 + 165 p \text{ mod } p^2 \\
x_4=s(1+r) \equiv 82 \text{ mod } p^2 \\
\end{array}
\]
For \(0\le m < 180\) we have \(u^0_m(r,s)\equiv 0\) mod \(p\) only when \(m
\in \{0,1,4,6\}\).
Applying \(x_i \rightarrow \frac{x_i^{p-1}-1}{p}\) gives
\[
\begin{array}{l}
y_1 \equiv 40 \text{ mod } p \\
y_2 \equiv 33 \text{ mod } p \\
y_3 \equiv 140 \text{ mod } p \\
y_4 \equiv 46 \text{ mod } p
\end{array}
\]
For \(m \in \{0,1,4,6\}\) the function \(u^1_m\)
\[
33\cdot 66^m+46 \cdot 82^m - 40 \cdot 140^m - 140 \cdot 9^m
\]
is nonzero modulo \(p\). Both Skolems lemmas can be applied and it is shown
that there is no
non-trivial symmetry but at order 6.
\section{The counter example to Fokas' conjecture}\label{Sec4}
\begin{theorem}
\label{fok}
There exists a 2-component equation with exactly two non-trivial symmetries.
\end{theorem}
\begin{proof}
The resultant of \(U_7\) and \(U_{11}\) has the following factor in common
with the
resultant of \(U_7\) and \(U_{29}\)
\[
(r^3-r-1)(r^3+r^2-1)(r^6+3r^5+5r^4+5r^3+5r^2+3r+1)
\]
In \(\Z/101\) the first factor has solution 20 and the third solution 52.
These can be lifted and
both Skolems lemmas can be applied. In this way it is proven that the set
\(\{ r,\frac{1}{r},s,\frac{1}{s} \}\)
that corresponds to the modulo \(101\) set \(S=\{20,96,52,68\}\) is no
solution set of a \(G_m\)-function when
\(m \not \in \{0,1,7,11,29\}\). If one of the other sets corresponding to
\(\{40,48,42,89\}\) or \(\{32,60,63,93\}\)
is a solution set of the function \(G_m\) for some \(m\) then their minimum
polynomials divides the
resultant of \(U_7\) with \(U_m\). That means that \(S\) is a solution set
of \(G_m\) as well, hence \(m\) equals
\(0,1,7,11\) or \(29\).
\end{proof}
We will compute the three equations and their symmetries explicitly. Each
root of \(r^3+r^2-1\)
is mapped to a different eigenvalue. We take \(\C[r]/(r^3+r^2-1)\) as our
coefficient field.
The eigenvalues of the systems will be
\[
1+r^7=2r^2 \text{ and } (1+r)^7=16r^2+28r+21
\]
Their quadratic part will be
\[
G_7[2r^2,16r^2+28r+21](\xi_1,\xi_2)
\]
divided by
\[
2(\xi_1-r\xi_2)(\xi_1-(r+r^2)\xi_2)(\xi_1^2+(1-r-r^2)\xi_1\xi_2+\xi_2^2)
\]
i.e.
\[
\frac{7}{2}(2r^2+4r+3)(\xi_1+\xi_2)(\xi_1^2+(2-r)\xi_1\xi_2+\xi_2^2)
\]
Our examples, written more compactly than in \cite{KS99}, look like
\[
\left.
\Bf{2r^2}{(16r^2+28r+21)}7{7(2r^2+4r+3)(v_3v_0+(3-r)v_2v_1)}\quad\right|
r^3+r^2-1=0.
\]
The symmetries can be calculated in the same way, leading to
\[
\begin{array}{ll}
u_t=& (-3r^2+r+2) u_{11} \\
&+ 11 \left( (14r^2+24r+18)v_7v_0 + (35r^2+57r+42)v_6v_1 \right.\\
&+ \left. (48r^2+70r+49)v_5v_2 + (51r^2+65r+42)v_4v_3 \right) \\
v_t=& (151r^2+265r+200) v_{11}
\end{array}
\]
and
\[
\begin{array}{ll}
u_t=& (-40r^2+9r+17) u_{29} \\
&+ 29\left( 30(1081r^2+1897r+1432)v_{25}v_0 \right. \\
&+ (311920r^2+547311r+413143)v_{24}v_1 \\
&+ (706832r+533441+403277r^2)v_{23}v_2 \\
&+ (449543r^2+782050r+589257)v_{22}v_3 \\
&+ (537572r+402545+317304r^2)/2v_{21}v_4 \\
&+ (1026233r^2+1635821r+1205570)v_{20}v_5 \\
&+ (1101516r+779787+787277r^2)/2v_{19}v_6 \\
&+ (2656229r+1710194+2393075r^2)v_{18}v_7 \\
&+ (3831912r^2+3208669r+1731205)v_{17}v_8 \\
&+ (6105788r^2+4007995r+1678107)v_{16}v_9 \\
&+ (4807604r+1421555+8899703r^2)v_{15}v_{10} \\
&+ (5263833r+11440843r^2+915604)v_{14}v_{11} \\
&+ \left. 3(1793035r+155000+4312473r^2)v_{13}v_{12} \right) \\
v_t=& (3761840r^2+6601569r+4983377) v_{29}
\end{array}
\]
\section{More symmetries}
We present the results of large computer calculations we did in MAPLE
\cite{Map91}.
We calculated the resultant of \(U_n\) and \(U_m\) for \(4\leq n \leq 10\) and
\(n+1 \leq m \leq n+150\). To obtain the systems with finitely many
symmetries only,
one has to filter out the integrable systems. How to find all integrable
\(n^{\text{th}}\)
order systems for fixed \(n\) if the quadratic part is \(v_0^2\) is
described in \cite{BSW99}.
This method we have extended to cover the case of an arbitrary quadratic
part. This will be
the subject of another paper. One finds common factors on orders
\(m=k(n-1),kn,(2k+1)n\) for
all \(k \in \mathbb{N}\).

To give an indication of the size of the expressions. The resultant of
\(U_{10}\) and \(U_{160}\)
has degree \(556\). The coefficients of \(r^n\) with \(244<n<312\) have 207
(decimal) digets.
The number of \(n^{\text{th}}\) order systems we have been calculating is
\begin{center}
\begin{tabular}{|c|c|c|c|c|c|c|c|c|c|}
\hline
 & & & & & & & & \\
n & 4 & 5 & 6 & 7 & 8 & 9 & 10 & 4--10 \\
 & & & & & & & & \\
 \hline
& & & & & & & & \\
\# & 2745 & 2701 & 5679 & 5644 & 8740 & 8839 & 11952 & 46300 \\
 & & & & & & & & \\
\hline
\end{tabular}
\end{center}
In the pictures on the next pages the positions of the roots of these
resultants in the complex
plane are plotted. As a fundamental domain the upper half unit circle is
chosen. The full pictures are
invariant under \(r\rightarrow \frac{1}{r}\) and \(r\rightarrow \bar{r}\).

\begin{figure}
\label{plot1}
\begin{center}
\includegraphics[width=0.5\textwidth, angle=270]{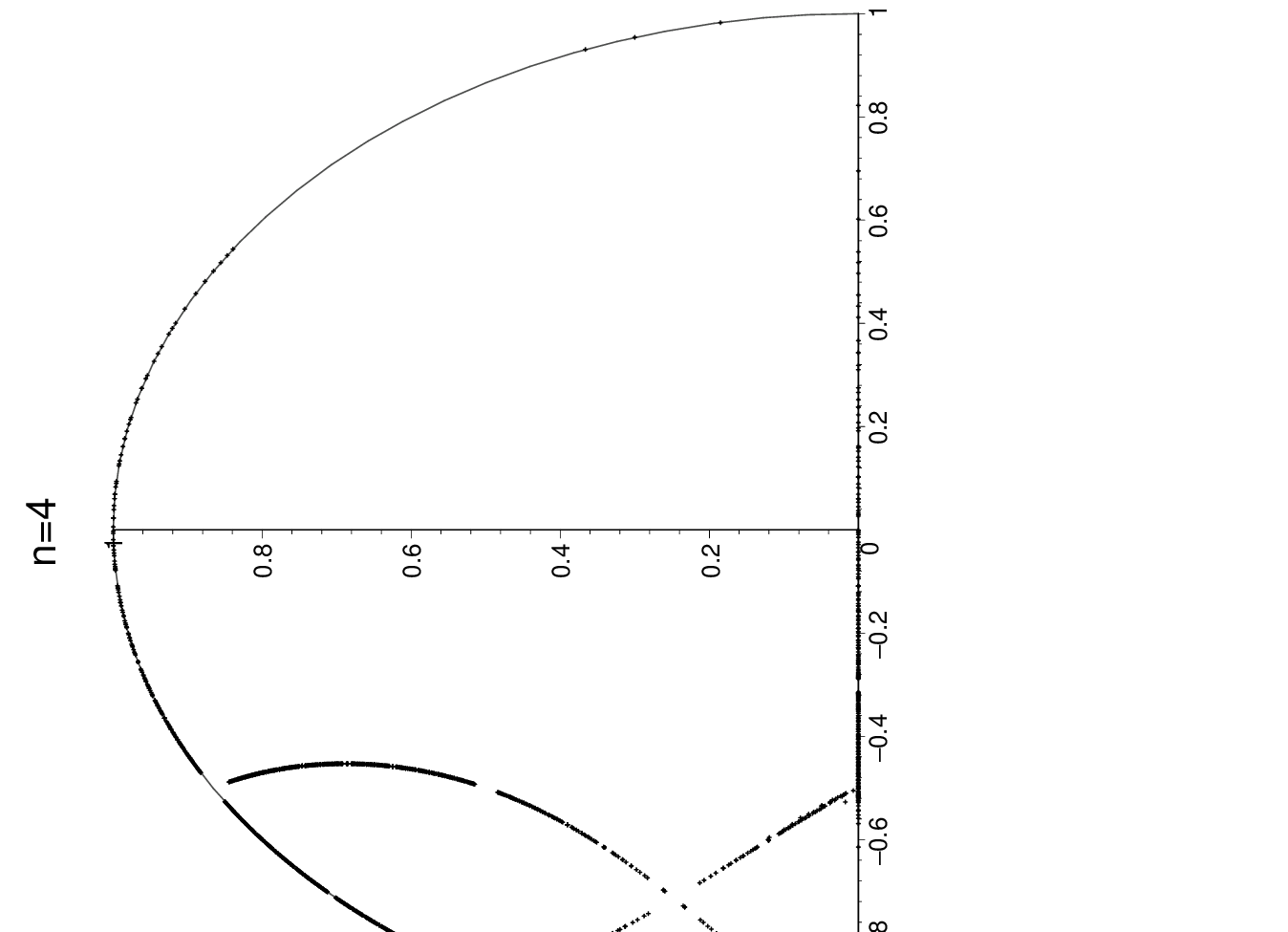}\\
\includegraphics[width=0.5\textwidth, angle=270]{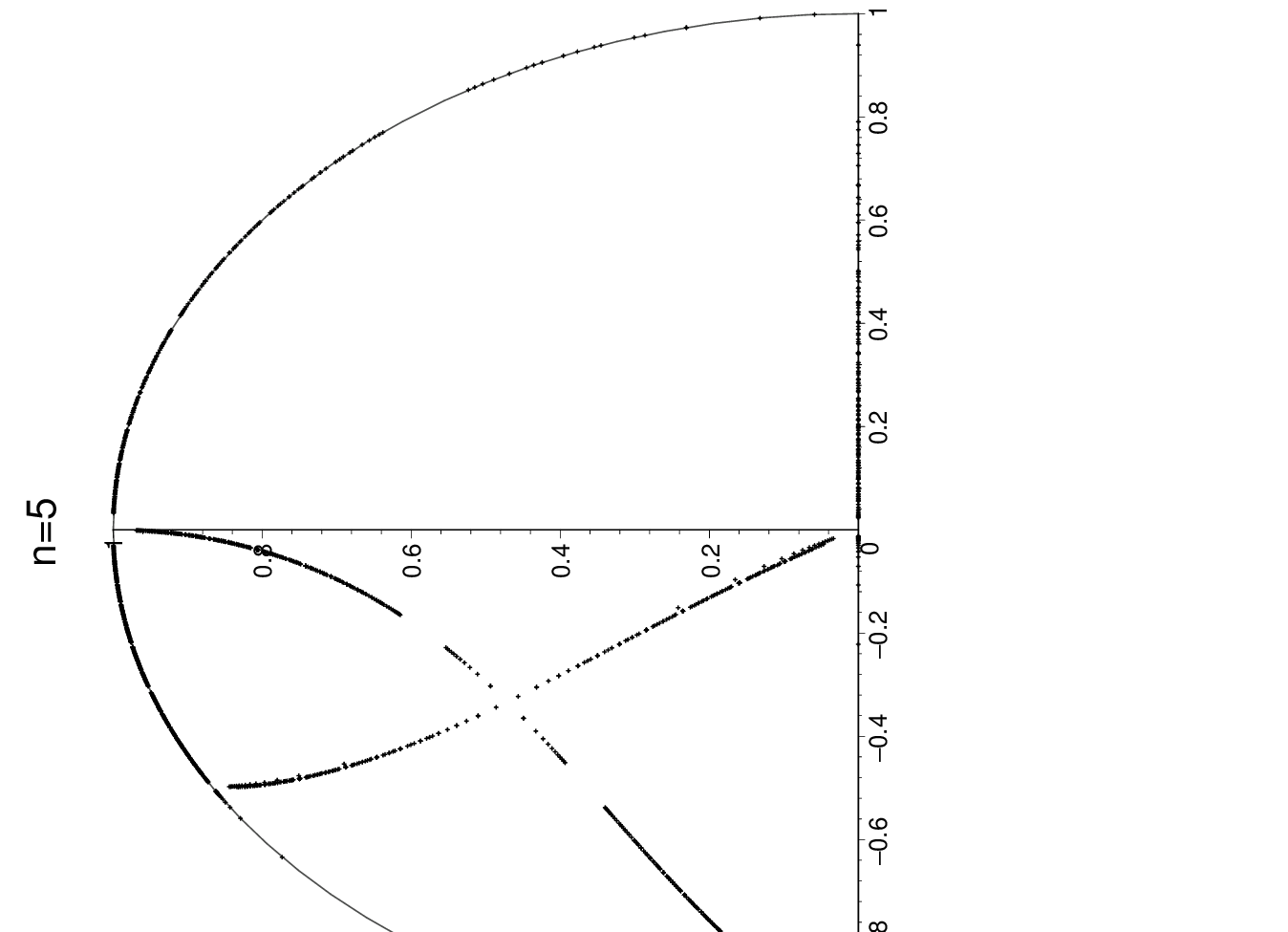}\\
\includegraphics[width=0.5\textwidth, angle=270]{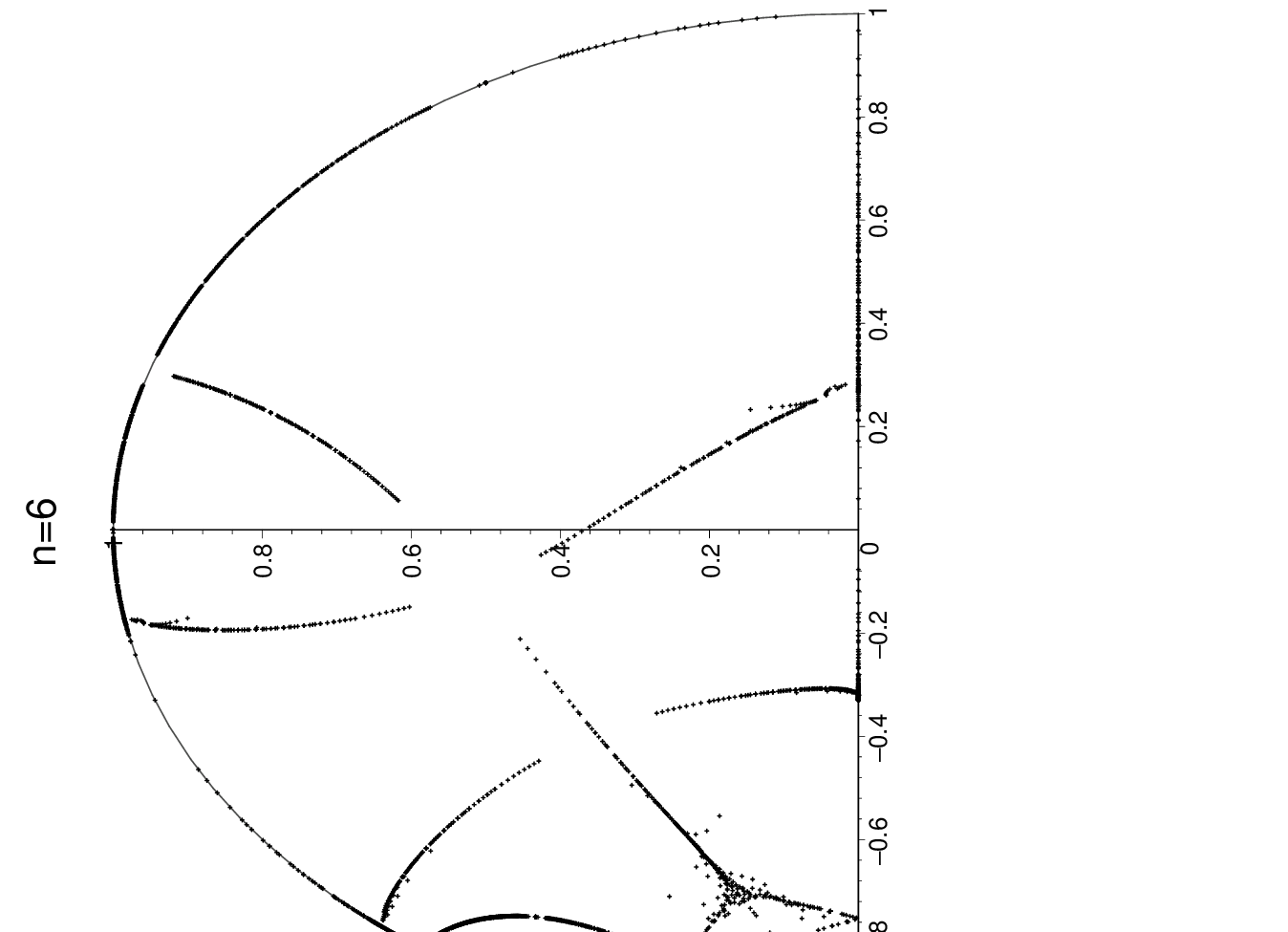}
\caption{Numerical approximations of the roots of the \(G\)--functions
corresponding
to almost integrable systems for order \(4,5\) and \(6\)}
\end{center}
\end{figure}

\begin{figure}
\label{plot2}
\begin{center}
\includegraphics[width=0.5\textwidth, angle=270]{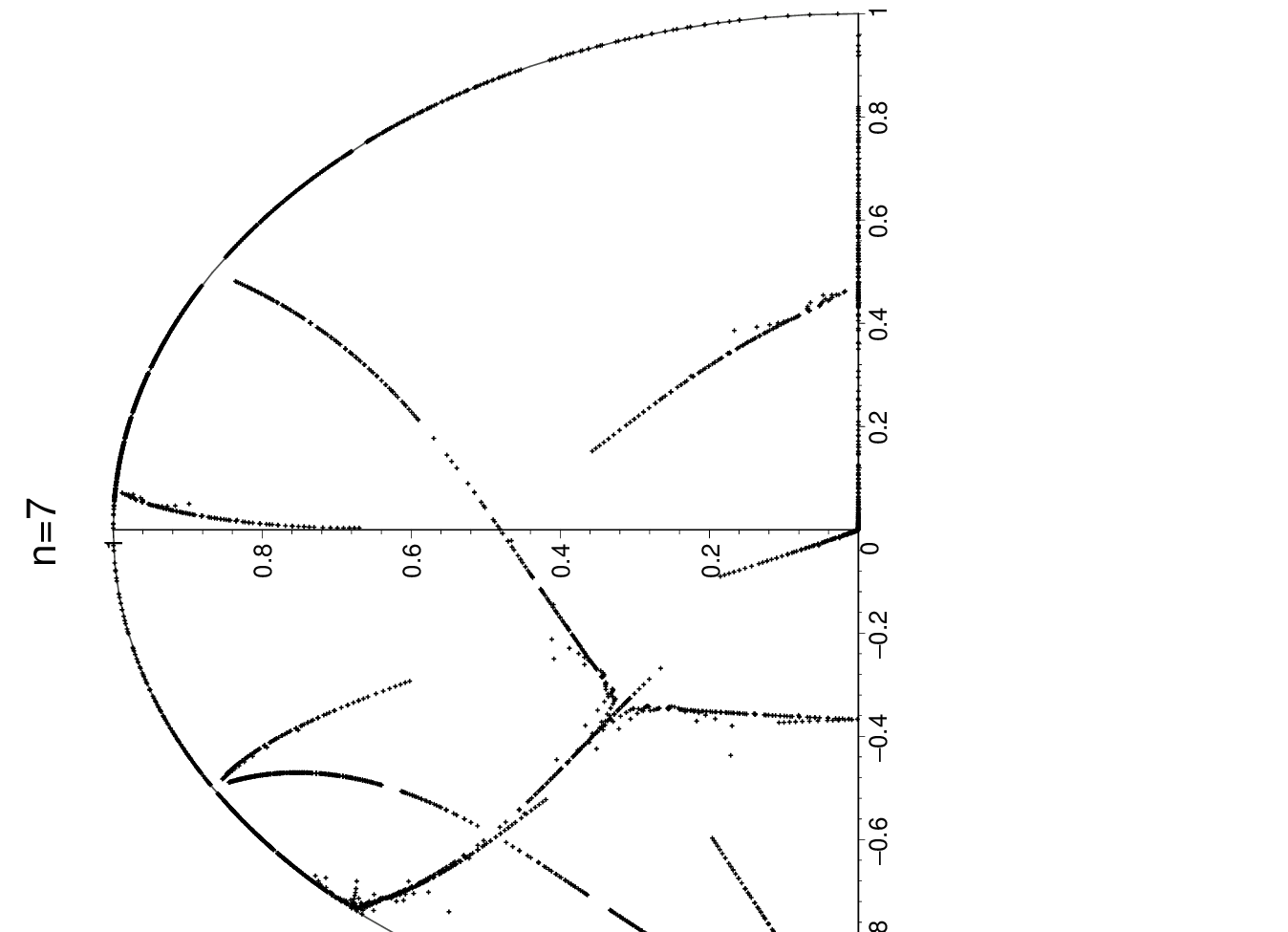}\\
\includegraphics[width=0.5\textwidth, angle=270]{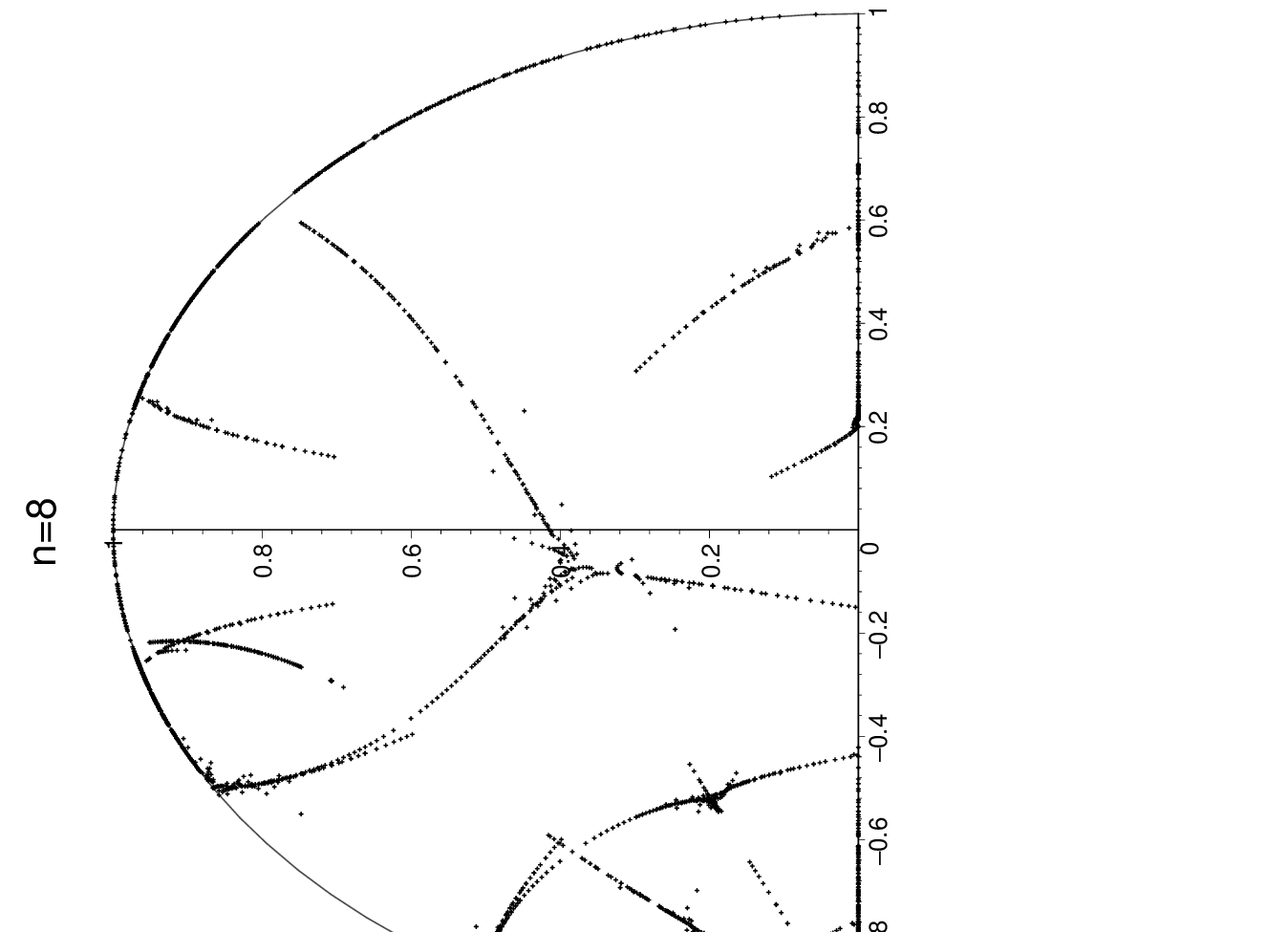}\\
\includegraphics[width=0.5\textwidth, angle=270]{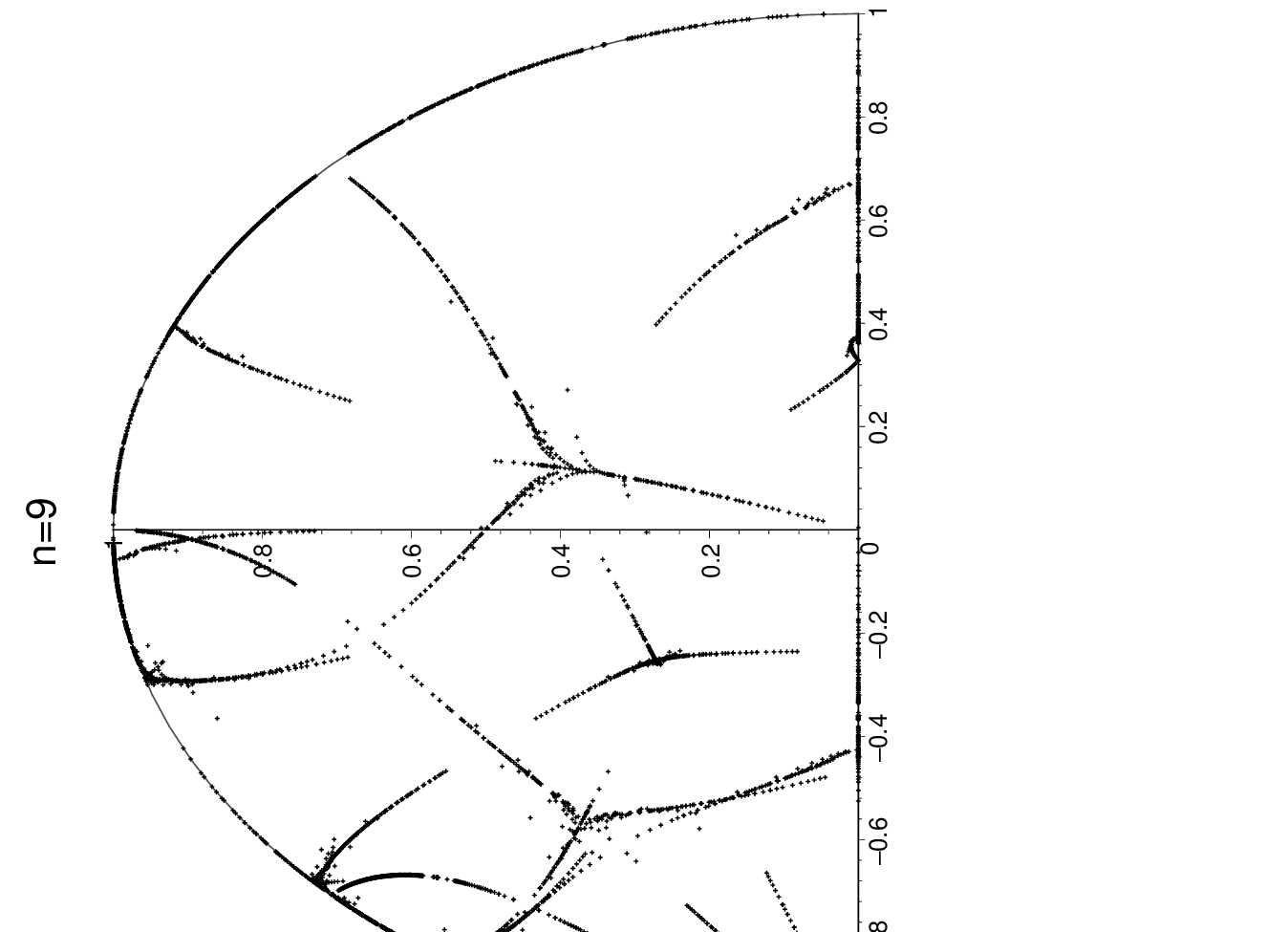}
\caption{Numerical approximations of the roots of the \(G\)--functions
corresponding
to almost integrable systems for order \(7, 8\) and \(9\)}
\end{center}
\end{figure}

\begin{figure}
\label{plot3}
\begin{center}
\includegraphics[width=0.5\textwidth, angle=270]{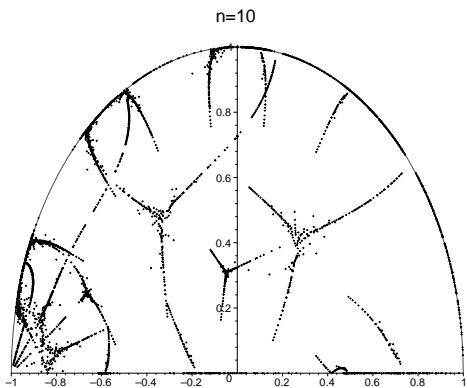}
\caption{Numerical approximations of the roots of the \(G\)--functions
corresponding
to almost integrable systems for order \(10\)}
\end{center}
\end{figure}

All these systems have at least one nontrivial symmetry. To answer the question
how many symmetries there exactly are we implemented the method of Skolem.
We made the following refinements.

\begin{itemize}
\item Most of the resultants we have calculated are irreducible. By the
argument
in the proof of theorem \ref{fok} it suffices to prove the statement for one
particular set of roots.
\item Sometimes it is much more efficient to use two pairs of roots. The
argument
goes as follows. The resultant of \(U_5\) and \(U_{19}\) contains the factor
\[
\begin{array}{ll}
f(r)=&r^{12}+4r^{11}+10r^{10}+19r^9+28r^8+34r^7 \\
&+37r^6+34r^5+28r^4+19r^3+10r^2+4r+1
\end{array}
\]
which is irreducible over \(\mathbb{Q}\) and splits into linear factors
over \(\mathbb{Z}_{509}^{\times}\).
The numbers \((264, 407)\) are a solution for \(U_m(r,s)\) when \(m \in
\{0, 1, 5, 19, 256, 414\}\).
The numbers \((267, 300)\) are a solution for \(U_m(r,s)\) when \(m \in
\{0, 1, 5, 19, 162, 254\}\).
By using both pairs we can apply lemma \ref{skol1} for all \(0 \leq m <
508\) but \(\{0, 1, 5, 19\}\),
for which we can use the lemma \ref{skol2}. The computer could not find any
prime such that the
normal procedure works, it has been busy for days to check all primes
\(p<8147\).
\end{itemize}
With these improvements we have been able to prove that all these system
have exactly one
non trivial symmetry, with the exeption of the seventh order systems with
two symmetries at
order 11 and 29.

The following MAPLE output can be used to verify the above statement for
\(n=7,29 \leq m \leq 37\).
\begin{verbatim}
prf29:=[101, {20, 52}],[97, {4, 32}]:
prf30:=[2531, {75, 871}]:
prf31:=[1021, {16, 42}]:
prf32:=[877, {226, 214}]:
prf33:=[601, {23, 409}]:
prf34:=[2857, {2457, 716}, {742, 391}]:
prf35:=[661, {401, 330}, {122, 245}]:
prf36:=[179, {17, 76}]:
prf37:=[233, {30, 56}, {20, 84}]:
\end{verbatim}
The sequence {\tt prf.m} contains the proofs for different factors of the
resultant of \(U_7\)
and \(U_m\). Each {\em proof} consists of a prime number \(p\) and one or
two sets of modulo \(p\) solutions
such that all conditions of Skolem are satisfied. The exceptions, where the
resultant has two factors, are
\[
(n,m)=(4,24),(4,28),(6,42),(7,8),(7,49),(8,56),(10,70)
\]
Three factors appear at \(n=7,m=11\) and four at \(n=7,m=29\).
\section{Conclusions}
We have shown that the existence of one or two generalized symmetries
of an evolution equation does not necessarily imply integrability.
We hope that this illustrates the use of p-adic and resultant methods
to this field and that these methods will be more widely applied.
With these results in mind this puts a burden of proof on
anyone claiming integrability (with respect to generalized symmetries).
We mention the successful use of number theoretic methods,
especially the Lech-Mahler theorem, in this respect, cf. \cite{BSW99,SW01a}.
These methods are not restricted to the special kind of systems
we study here, but they are applicable to any polynomial system,
in principle.

\label{lastpage}

\end{document}